\documentclass[aps,prl,twocolumn,superscriptaddress]{revtex4}

\usepackage{amsmath,amssymb}
\usepackage{slashed}	

\usepackage{graphicx}
\usepackage{fancybox,color}

\usepackage[active]{srcltx} 
\usepackage[utf8]{inputenc} 
\usepackage[T1]{fontenc} 

\usepackage{enumitem}
\setitemize[1]{itemsep=0.5em,parsep=0em}

\usepackage[font=footnotesize,labelfont=bf, justification=centerlast]{caption}

\usepackage{xspace}

\usepackage[english]{babel}
\usepackage{amsfonts}
\usepackage{lmodern}

\def\xnewpage{} 

\def\off#1{}

\newcommand{\mybox}[2][blue]{{\color{#1}\fbox{\color{black} #2}}}
\newlength{\myboxL}

\definecolor{mygray}{gray}{0.5}

\newcommand{\xcite}[1][...]{\mybox{cite #1}}

\def\RS#1{\red{#1}}

\def\ra{\rightarrow}

\def\hc{\text{h.c.}}

\def\>{\rangle}
\def\<{\langle}

\renewcommand{\Re}{\text{Re}\,}
\renewcommand{\Im}{\text{Im}\,}
\def\v{\mathbf}
\def\p{\partial}
\def\h{\hat}

\def\s{{}^\dagger}
\def\sph{{}^{\phantom{\dagger}}}

\def\H{{\cal H}}
\def\F{{\cal F}}

\def\N{{\cal N}}
\def\D{\Delta}

\def\la{\lambda}
\def\d{\delta}

\newcommand{\bea}{\begin{eqnarray}}
\newcommand{\ea}{\end{eqnarray}}
\newcommand{\eea}{\end{eqnarray}}

\newcommand{\iu}{\mathrm{i}}
\newcommand{\dd}{\mathrm{d}}

\definecolor{Green}{RGB}{50,113,34}
\definecolor{Red}{RGB}{243,23,28}
\definecolor{Blue}{RGB}{32,142,211}
\definecolor{Purple}{RGB}{186,15,211}

\renewcommand{\section}[1]{\paragraph{#1.}}
\renewcommand{\subsection}[1]{}

\newif\iffinal
  \finaltrue		

\iffinal
  \def\finaloff#1{}
  \renewcommand{\mybox}[2][blue]{}
  \renewcommand{\xcite}[1][...]{[...]}
  
  \def\RS#1{{#1}}
\else
  \def\finaloff#1{#1}
\fi

\begin{document}

\title{Quantum simulation of spontaneous pair creation in 2D optical lattices}

\author{Leonhard Klar}
\affiliation{Fakult\"at f\"ur Physik, Universit\"at Duisburg-Essen, Lotharstra{\ss}e 1, Duisburg 47057, Germany,}

\author{Nikodem Szpak}
\email{nikodem.szpak@uni-due.de}
\affiliation{Fakult\"at f\"ur Physik, Universit\"at Duisburg-Essen, Lotharstra{\ss}e 1, Duisburg 47057, Germany,}

\author{Ralf Schützhold}
\affiliation{Fakult\"at f\"ur Physik, Universit\"at Duisburg-Essen, Lotharstra{\ss}e 1, Duisburg 47057, Germany,}
\affiliation{Helmholtz-Zentrum Dresden-Rossendorf, Bautzner Landstra{\ss}e 400, 01328 Dresden, Germany,}
\affiliation{Institut f\"ur Theoretische Physik, Technische Universit\"at Dresden, 01062 Dresden, Germany.}

\date{\today}

\begin{abstract}
  One of the fundamental predictions of Quantum Electrodynamics (QED) is the spontaneous 
  creation of particle--antiparticle pairs from vacuum in presence of a very strong electric field. 
  Under these extreme conditions a strongly bound state 
  can fetch an otherwise unobservable electron from the Dirac sea, 
  leaving behind a hole representing a positron.
  Although generally known for many decades, the effect has not yet been demonstrated experimentally. 
  We propose an analogue model of the quantum Dirac field, realized by ultra--cold 
  fermionic atoms in an optical lattice, aiming at an experimental simulation of this 
  intriguing non--perturbative phenomenon. 
  %
  %
  %
  Numerical simulations demonstrate the effect of spontaneous pair creation in the optical analogue system, in qualitative agreement with QED: 
  \RS{in the adiabatic regime the vacuum can be destabilized only by supercritical fields exceeding a critical threshold.}
\end{abstract}

\maketitle

\section{Introduction}

The relativistic Dirac theory of fermions describes electrons ($e^-$) and positrons ($e^+$).
These correspond to the positive ($E\geq mc^2$) and negative ($E\leq-mc^2$) energy solutions of the Dirac equation \cite{Thaller}.
A localized electric potential $V(\v x)$ (e.g. of an atomic nucleus) can introduce bound states in the spectral gap $(-mc^2, +mc^2)$.
The electronic bound states can be shifted by the strength of the negative potential from the upper to the lower continuum edge and interesting physics begins when the negative continuum is reached.
If such a bound state dives into the negative continuum it turns into a resonance 
and decays in time \cite{Rafelski+Mueller+Greiner-ChargedVac-Overcrit, GrMuRa, NS-PhD, NS-QED-JPhysA, Pickl+Duerr-AdiabSpPCr}.
After it ``dives out'' again 
some part of the wave function stays mixed with the continuum. 
The adiabatic character of the process admits the interpretation 
of a particle being slowly pulled from the otherwise unobservable \textit{Dirac sea} 
while the resulting hole in the \textit{sea} appears as an antiparticle.
In contrast to the standard dynamical pair production, which vanishes in the adiabatic limit,
this type of pair production process, related to vacuum decay, survives the limit and is therefore called \textit{spontaneous}.

The spontaneous creation of true electron--positron pairs,
a fundamental prediction of QED  \cite{GrMuRa},  
has not yet been confirmed experimentally as the generation of a sufficiently strong
electric field to overcome the gap of $2\, m_e c^2 \approx 1$~MeV still presents a challenge \cite{Rafelski2017-ProbingVacuum}.
Fortunately, a distant area of low energy physics provides help: 
ultra--cold atoms moving in an optical lattice offer an analogue model for the strong field QED.
The fermionic vacuum can be replaced by the Fermi level for given average filling
and its quantum excitations, acting as quasi--particles, can mimic the behavior of elementary particles.
Since the mass gap in the analog model can be tuned freely, it is much easier to
create conditions necessary for the spontaneous pair creation.
This motivates our proposal for a quantum simulator in which excitations of
ultra--cold atoms moving in an optical lattice will represent quasi--relativistic particles
and antiparticles (holes) satisfying a discretized version of the Dirac
equation together with the fermionic anti--commutation relations.
By mapping onto each other the second quantized many particle Hamiltonians, we are able to construct an
analogue of the spontaneous pair creation which can be realized in a table--top experiment
\cite{Bloch-OptLat, MGreiner+Esslinger+Haensch+Bloch-SuperfluidMottOptLat, Damski+Lewenstein+Sanpera-OpticalLattices-Review, Esslinger-FermiHubbard-Review, Krutitsky-BoseHubbard-Review, Bloch+Dalibard-OptLat-Review, Yukalov-OptLat-Review}.

Our proposal aims at the simulation of the non--perturbative effect laying in the heart of the spontaneous pair creation via an adiabatic destabilization of the ground state.
%
It also opens the way for simulation of many--body effects such as particle--hole creation and annihilation, vacuum polarization or the impact of interactions, 
a subject which is difficult to address theoretically or even computationally.
In the context of QED, it has been argued that localized supercritical electric fields will not exist 
due to the strong screening effects caused by virtual particle--antiparticle pairs.
This non--linear reaction of the quantum vacuum to its strong perturbation has been hotly debated in the context of QED \cite{GrMuRa, Rafelski2017-ProbingVacuum} and graphene \cite{Pereira+CastroNeto-Supercritical-Graphene, Katsnelson+Levitov-VacuumPol-Supercritical-Graphene} but not fully clarified. 
Further development of quantum simulators in this direction can shed new light on this outstanding problem.
There exist proposals for the first 
\cite{Dirac-photons, Witthaut+Weitz-DoublePeriodicOptPot}
and second--quantized Dirac Hamiltonian
\cite{QSim-Dirac-HexLattice, Dirac+Interaction-OptLat, 
MasslessDirac-SqLattice, Dirac-StaggeredMagnField, Goldman+Lewenstein-Dirac-by-SU2, Lewenstein-Dirac-CurvedST, FieldTheo-QSim-In-OptLat, StrongB-Staggered-OptLat}
but they consider scenarios which are more involved than the setup
discussed here or aim at different models and effects.

In our previous works \cite{NS+RS-BiOptLat-Lett, NS+RS-BiOptLat}, some of us studied a similar question in one spatial dimension. 
Although the realization of the quantum optical setup is simpler in 1D,
the character of the bound state diving into the Fermi sea and thus of the pair creation process are slightly different \cite{NS+RS-BiOptLat-Lett, NS+RS-BiOptLat}.
The 2D system is in this respect more similar to the 3D case but simpler to realize.

\xnewpage
\section{Simulation of Dirac fermions in optical lattice}

\subsection{The Dirac Equation in 2+1D}

The Dirac equation,
  $ \gamma^{\mu}\left(\iu\p_{\mu}-qA_{\mu}\right)\Psi-M\Psi=0$
(in natural units $c=\hbar=1$),
describes relativistic fermions with mass $M$ and charge $q$. 
$A_\mu$ is the electromagnetic vector potential
and $\gamma^\mu$ are the Dirac matrices satisfying the Clifford algebra 
$\left\lbrace\gamma^{\mu},\gamma^{\nu}\right\rbrace=2\,\eta^{\mu\nu}$
with $\eta^{\mu\nu}$ being the Minkowski metric.
In two spatial dimensions, $\gamma^\mu$
can be represented by the Pauli matrices $\sigma_i$ such that
$\gamma^0=\sigma_3$ and $\gamma^0\gamma^i=-\sigma_i$
for $i=1, 2$.
In the following, we consider only the electric field and choose the gauge
$q A_{\mu}=\left(\Phi,0,0\right)$
with time and space dependent $\Phi(t,\v x)$.
This gives an evolution equation,  
$  \iu \p_t\Psi=\left(\iu\sigma_1\p_x+\iu\sigma_2\p_y+\sigma_3M+\Phi\right)\Psi $,
with the corresponding Hamiltonian
\begin{align} \label{eq:Dirac-Ham}
  H&=\int\dd^2x\,\Psi^{\dagger}\left(\iu\sigma_1\p_x+\iu\sigma_2\p_y+\sigma_3M+\Phi\right)\Psi\,.
\end{align}

\subsection{Discretization on 2D square lattice}

In order to obtain an analogue model for the Dirac field using an optical lattice system,
a standard space discretization procedure is performed \cite{NS+RS-BiOptLat-Lett, NS+RS-BiOptLat}. 
%
The discrete, two--component wavefunction $\psi_{m,n}$ will be defined on the two dimensional square lattice $\mathbb{Z}^2$ with lattice constant $l$ whereas the time variable will remain continuous.

\subsection{Many particle picture} 

\begin{figure}[ht]
  \begin{minipage}[t][][b]{0.45\linewidth}
    \includegraphics[height=3.6cm]{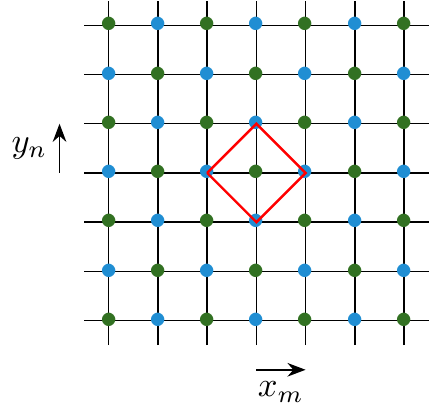}
  \end{minipage}
  \hskip 0.2cm
  \begin{minipage}[t][][b]{0.45\linewidth}
    \vskip -1.5mm
    \includegraphics[height=3.3cm]{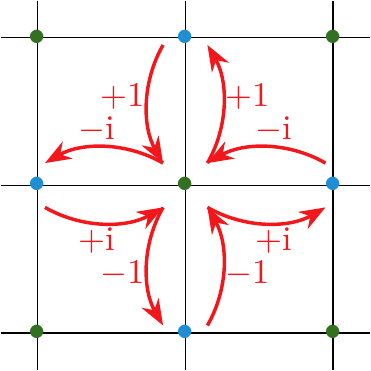}
  \end{minipage}
  \caption{Left: Square lattice decomposed into sublattices $A$ (green) and $B$ (blue) with supercell marked.
    Right: Tunneling amplitudes on the square lattice.
  }
  \label{fig:lattice}
\end{figure}

For many particle description, including particle creation and annihilation processes, 
the second quantization is introduced
in which the discretized field operator $\h\psi_{n,m}$ satisfies the fermionic anti--commutation relations on the lattice
$ \left\lbrace\hat{\psi}^{\alpha}_{m,n},\hat{\psi}^{\alpha'\dagger}_{m',n'}\right\rbrace = \d_{m,m'}\d_{n,n'}\d^{\alpha,\alpha'}, 
  \left\lbrace\hat{\psi}^{\alpha}_{m,n},\hat{\psi}^{\alpha'}_{m',n'}\right\rbrace =
  \left\lbrace\hat{\psi}^{\alpha\dagger}_{m,n},\hat{\psi}^{\alpha'\dagger}_{m',n'}\right\rbrace\ = 0$,
where $\alpha=1, 2$ refers to the spinor components of $\h\psi_{n,m}$.
Representing the spinor as
$
  \hat{\psi}_{m,n} = (
  \hat{a}_{m,n}, \hat{b}_{m,n}
  )^T
$
%
the second quantized discretized Hamiltonian 
takes the form
\begin{align} \label{eq:discr-Ham}
  \hat{H} &= \sum_{m,n} \Big[ \frac{1}{2l} \Big( \iu\hat{b}^{\dagger}_{m+1,n}\hat{a}\sph_{m,n} - \iu\hat{b}^{\dagger}_{m-1,n}\hat{a}\sph_{m,n} \nonumber \\
  & \phantom{= \sum_{m,n} \Big[ \frac{1}{2l}} + \hat{b}^{\dagger}_{m,n+1}\hat{a}\sph_{m,n} -\hat{b}^{\dagger}_{m,n-1}\hat{a}\sph_{m,n} + \hc \Big) \nonumber \\
  +& \left(\phi_{m,n}+M\right)\hat{a}^{\dagger}_{m,n}\hat{a}\sph_{m,n} 
  + \left(\phi_{m,n} -M\right)\hat{b}^{\dagger}_{m,n}\hat{b}\sph_{m,n} \Big].
\end{align}
The first two lines describe tunneling (also referred to as \textit{hopping}) between the neighboring lattice sites 
while the last line contains the potential and mass terms. 
This Hamiltonian splits into two separate parts, each having a staggered structure and involving only a half of all the operators $\hat{a}_{m,n}$ and $\hat{b}_{m,n}$ located at even (A) and odd (B) sites, respectively, as illustrated in Fig. \ref{fig:lattice}{, left}.
Hence, the lattice splits into two identical copies containing the same physics,
so it will be sufficient to consider only one of them%
\footnote{This eliminates half of the fermion doubling problem which typically occurs on the lattice \cite{Susskind-FermionDoubling}. 
In 2D, $2^d=4$ copies arise in total so two still remain in the system.}.
The supercell contains now two lattice sites, one in each of the sublattices A and B, and is diamond shaped. 
The structure of the complex tunneling amplitudes is shown in Fig. \ref{fig:lattice}{, right}.

\subsection{Free spectrum}

\begin{figure}[ht]
  \begin{minipage}[c][][b]{0.49\linewidth}
    \includegraphics[width=\linewidth]{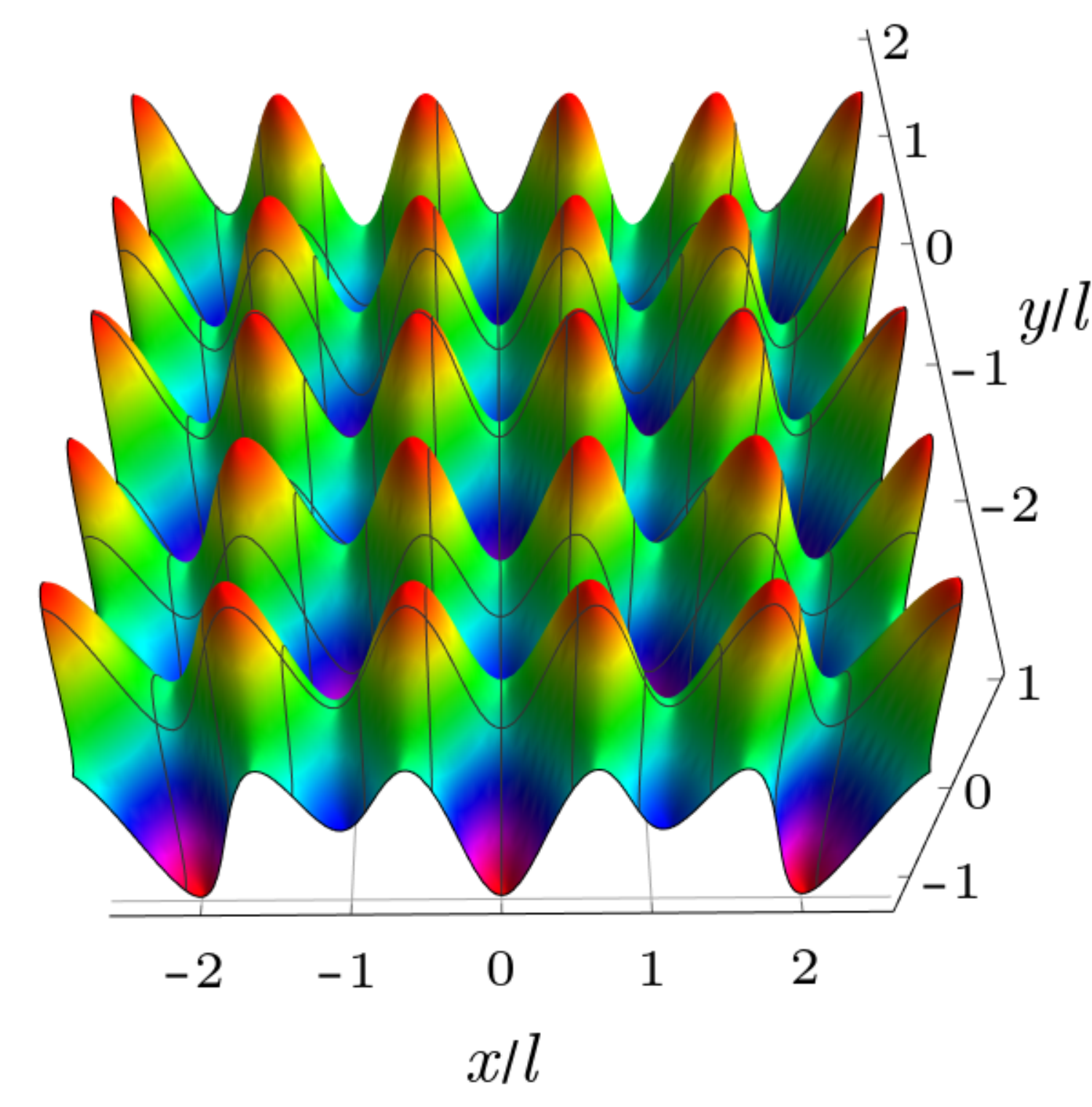}
  \end{minipage}
  \begin{minipage}[c][][b]{0.49\linewidth}
    \includegraphics[width=\linewidth]{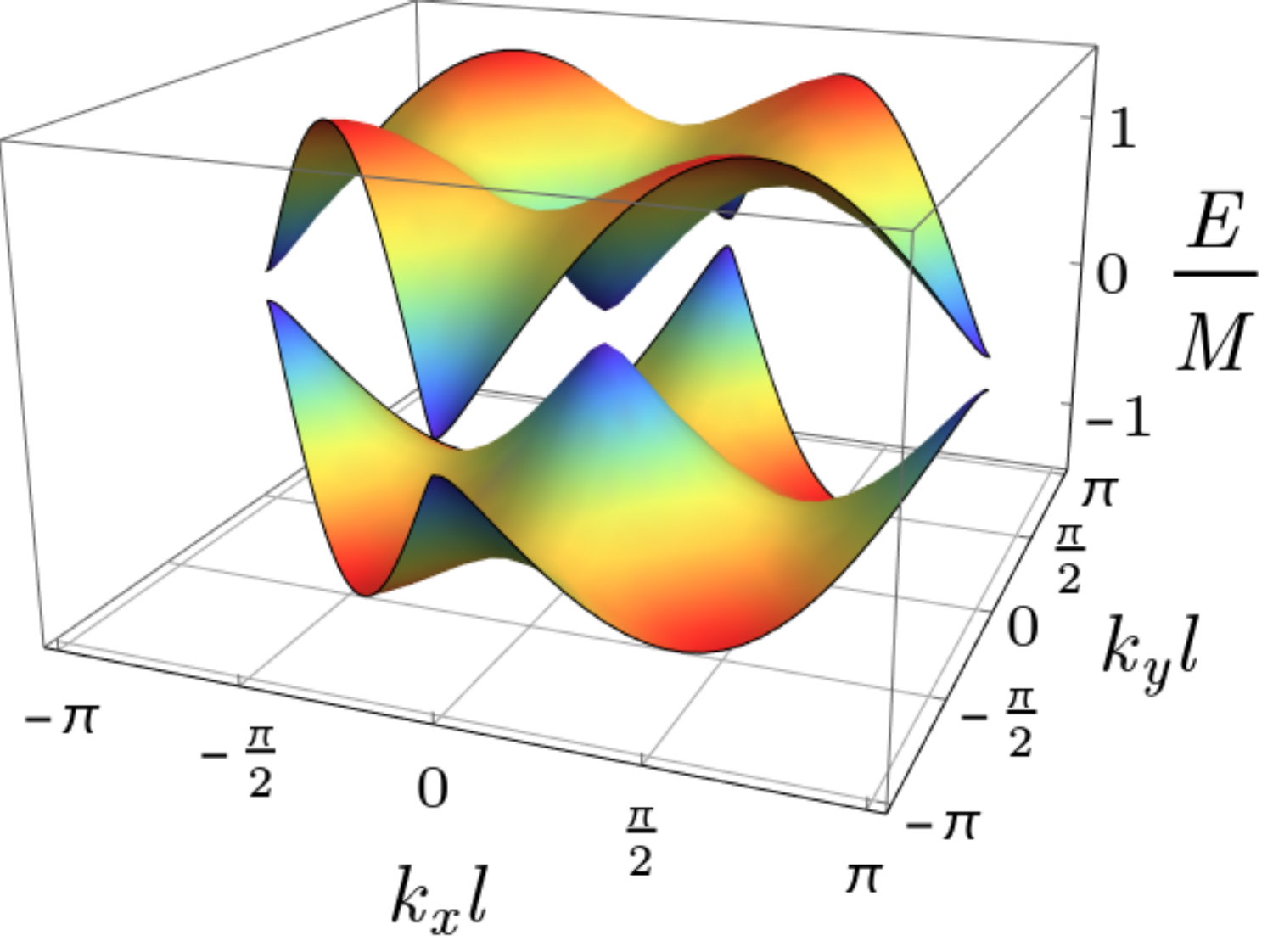}
  \end{minipage}
  \caption{2D bichromatic optical potential (left)
  and the dispersion relation $E(\mathbf{k})$ in the first Brillouin zone (right)} 
  \label{fig:disprel}
\end{figure}

If the external potential $\Phi$ is set to zero 
the dispersion relation $E(\mathbf{k})$ can be obtained analytically. 
Here, $\v k$ is the quasi--momentum defined in the first Brillouin zone 
$BZ = \{ \v k: |k_x| + |k_y| \leq \frac{\pi}{l}\}$ 
which is also diamond shaped.
By performing the Fourier transform from the lattice to the Brillouin zone and diagonalization,
the Hamiltonian \eqref{eq:discr-Ham} can be written as 
$  \tilde H = \int \dd^2k\, E(\v k)\, [ \h c^\dagger(\v k) \h c(\v k) - \h d^\dagger(\v k) \h d(\v k) ] $ 
%
with $\h c(\v k)$ and $\h d(\v k)$ referring to particles and antiparticles
and the dispersion relation
$  E(\v k) = \pm\sqrt{M^2 + \left(\sin^2k_xl+\sin^2k_yl\right)/l^2}$.
%
The spectrum consists of two bands separated by a gap of $2 M$ (cf. Fig. \ref{fig:disprel}{, right})
and can be realized in a bichromatic optical lattice 
\cite{Weitz-DoublePeriodicOptPot, Witthaut+Weitz-DoublePeriodicOptPot}.
It has two (inequivalent) hyperboloidal points $\v k_i$  where $E(\v k_i)\approx \pm M$, one at the center $\v k_0 = (0,0)$ and one at each corner of the Brillouin zone, $\v k_1 = (\pm\pi/l, 0)$ or $\v k_1 = (0, \pm\pi/l)$ (all four corners are equivalent). 
There, the dispersion relation approximates the relativistic one, $E \approx \pm \sqrt{M^2 + \d \v k^2}$ with $\d\v k = \v k - \v k_i$.
For $M=0$, these two points become gapless conical Dirac points with $E \approx \pm |\d\v k|$.
Referred to as pseudospin states, they are a consequence of the fermion doubling on the lattice \cite{Susskind-FermionDoubling}.

\subsection{Spectrum in presence of a binding potential and supercriticality}

\begin{figure}[ht]
  \includegraphics[width=0.48\linewidth]{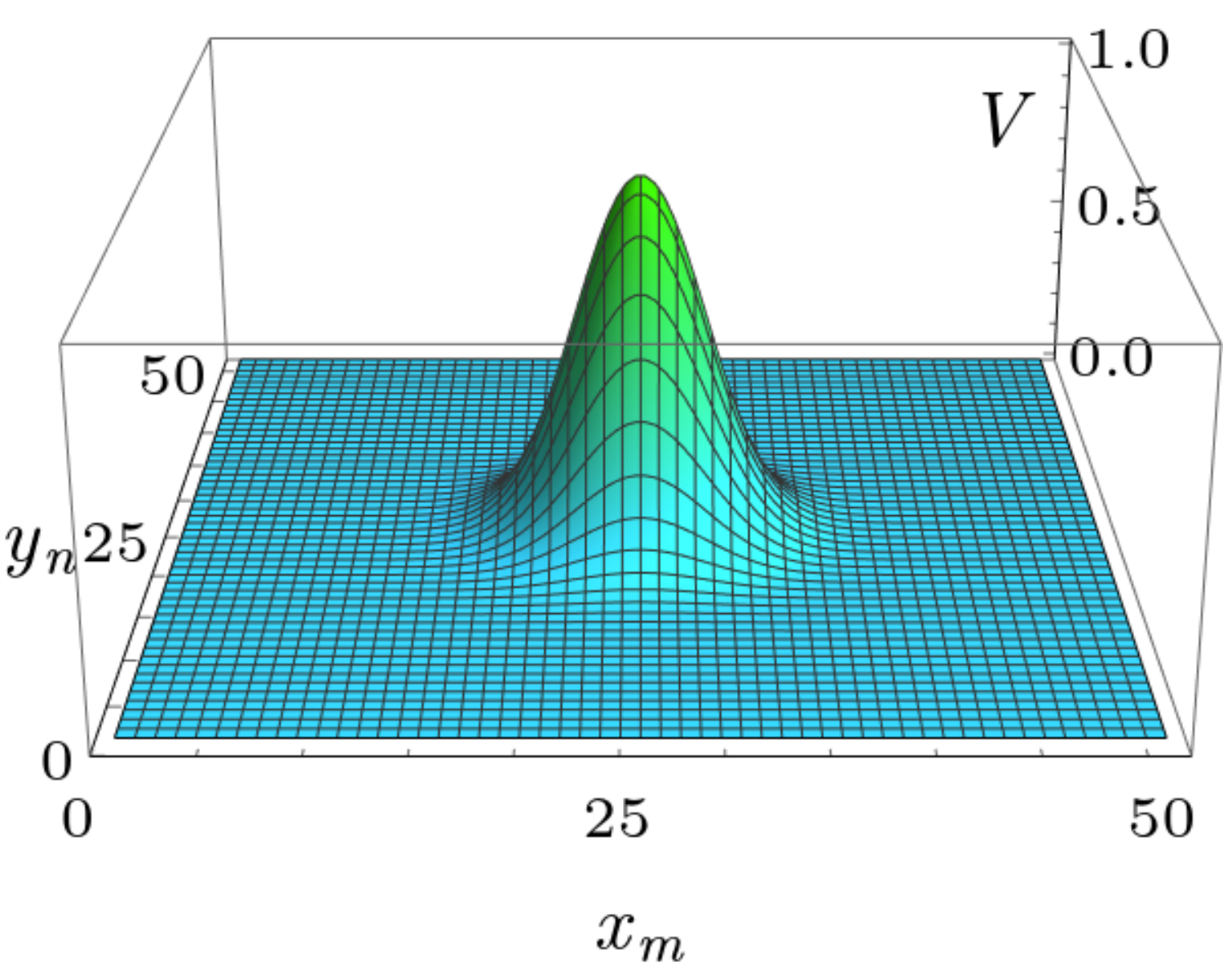}
  \hfill
  \includegraphics[width=0.48\linewidth]{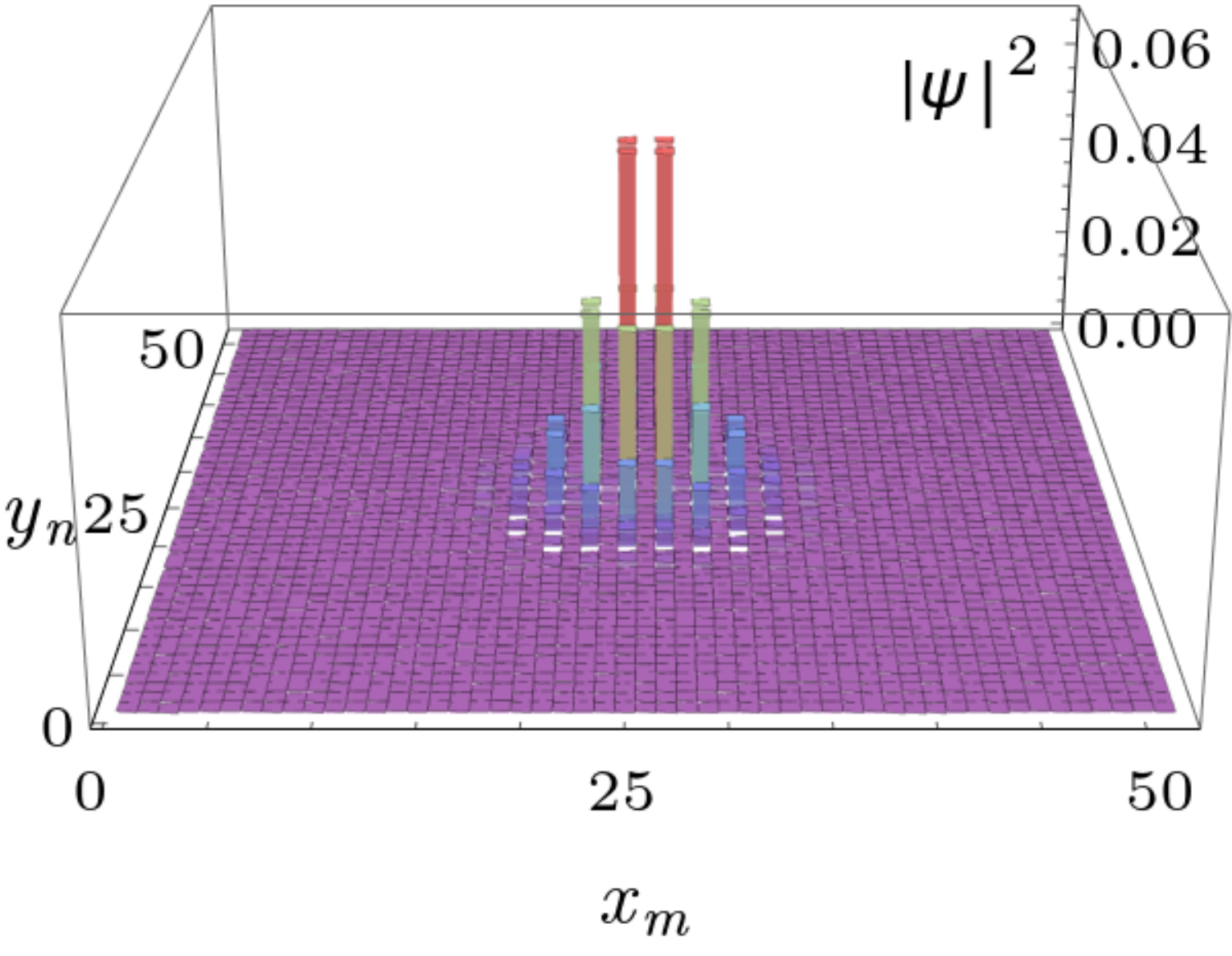}
  \caption{Left: Gaussian laser beam potential $V(\v x)$ with width $\sigma = 5\, l$. 
    Right: Strongly localized bound state with $E= - 0.87 M$. 
  }
  \label{fig:potential-exp}
\end{figure}

Now, we consider a family of strongly localized potentials, $\Phi_\la(\v x) = -\la V(\v x)$. We choose the Gaussian shape, $V(\v x) = V_0 \exp(-\v x^{\,2}/\sigma^2)$, cf. Fig. \ref{fig:potential-exp}{, left}, motivated by the optical lattice setup with an additional perpendicular narrow laser beam \cite{Bloch-OptLat}.
A negative potential, with $\la > 0$, will be binding for quasi--particles on the lattice. 
A typical bound state form is presented in Fig. \ref{fig:potential-exp}{, right}.

\begin{figure}[ht]
  \hskip -0.1cm
  \includegraphics[width=0.51\linewidth]{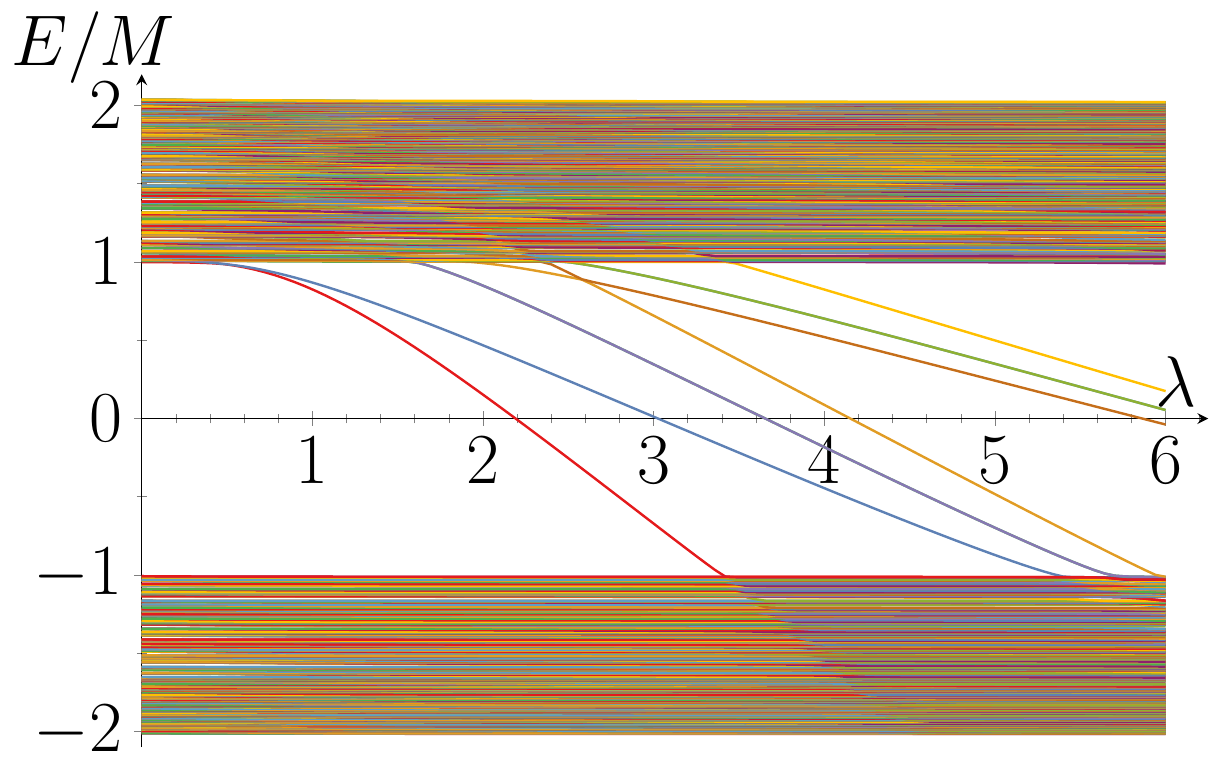}
  \hskip -0.2cm
  \includegraphics[width=0.51\linewidth]{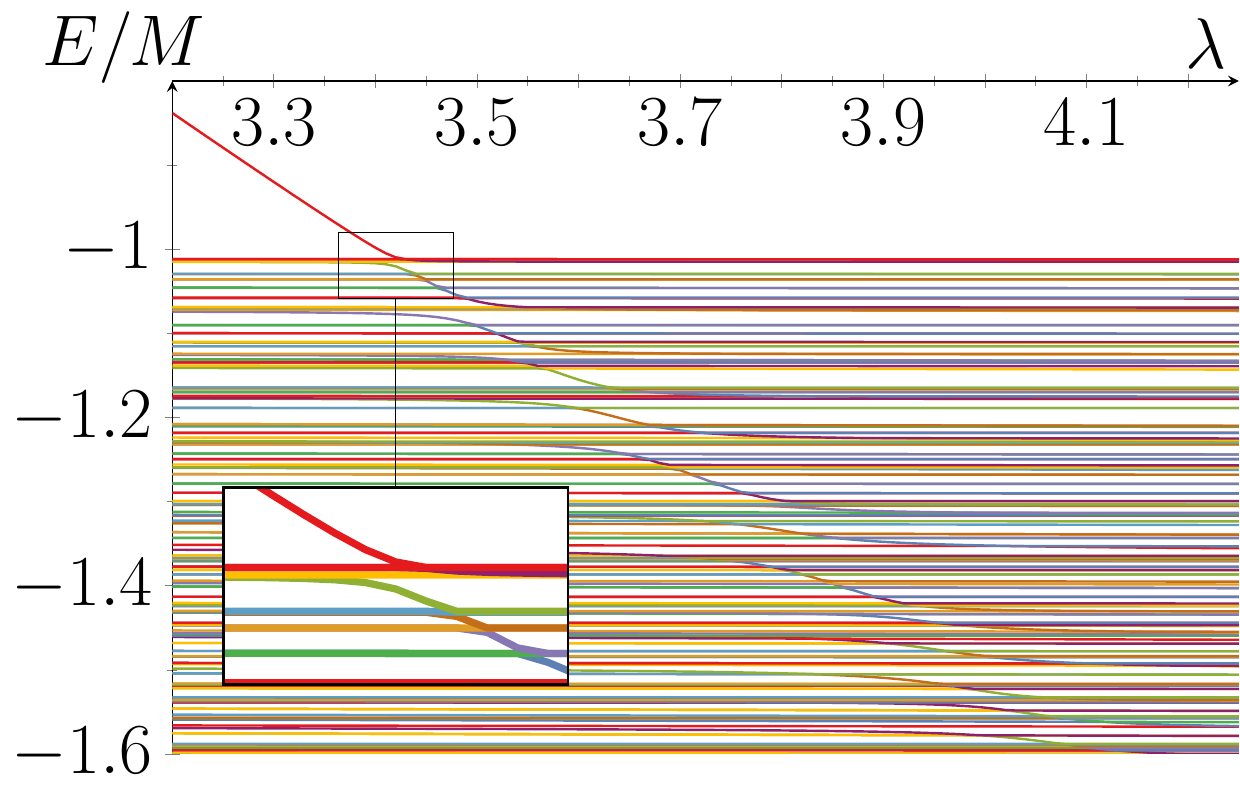}
  \caption{Left: Spectrum of the Hamiltonian $H(\la)$ as a function of the potential amplitude $\la$.
    Right: Zoom in the supercritical phase showing avoided crossings between the dived bound state and the discretized continuum.
  }
  \label{fig:spectrum-lambda}
\end{figure}

Increasing the value of $\la$ will cause the bound state energies to move monotonically down until the boundary of the negative continuum at $E = -M$ is reached for some critical $\la_{cr}$. In the continuous Dirac equation, the bound state turns into a resonance for supercritical $\la > \la_{cr}$. 
In the finite lattice model, the continuous spectrum is naturally discretized by the imposed boundary conditions.  The density of states in the pseudo--continuum is related to the total number of lattice sites and increases with the lattice size, thus approaching theoretically the continuum limit for infinitely large lattices.
In our system, the deepest bound state can cross the boundary at $E=-M$ and move further down producing either crossings or avoided crossings with the pseudo--continuum states. 
In general, for each value of $\lambda$ there exist simultaneously several bound states and all move monotonically down in the energy scale with increasing $\lambda$ (cf. Fig. \ref{fig:spectrum-lambda}{, left}).
A dived bound state marks its path through the continuum by a series of aligned avoided crossings (cf. Fig. \ref{fig:spectrum-lambda}{, right}).
%


\subsection{Evolution in time--dependent potential}

\begin{figure}[ht]
  \includegraphics[width=0.8\linewidth]{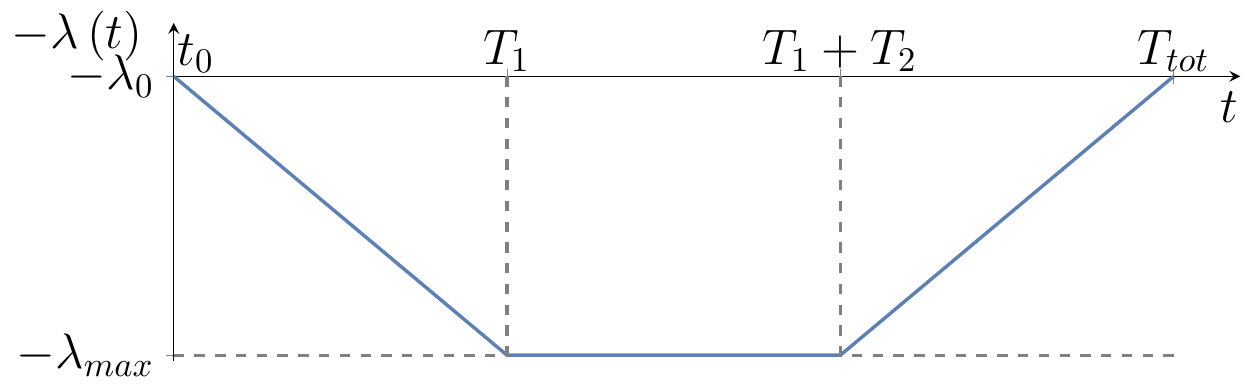} \\[0.5em]
  \includegraphics[width=0.8\linewidth]{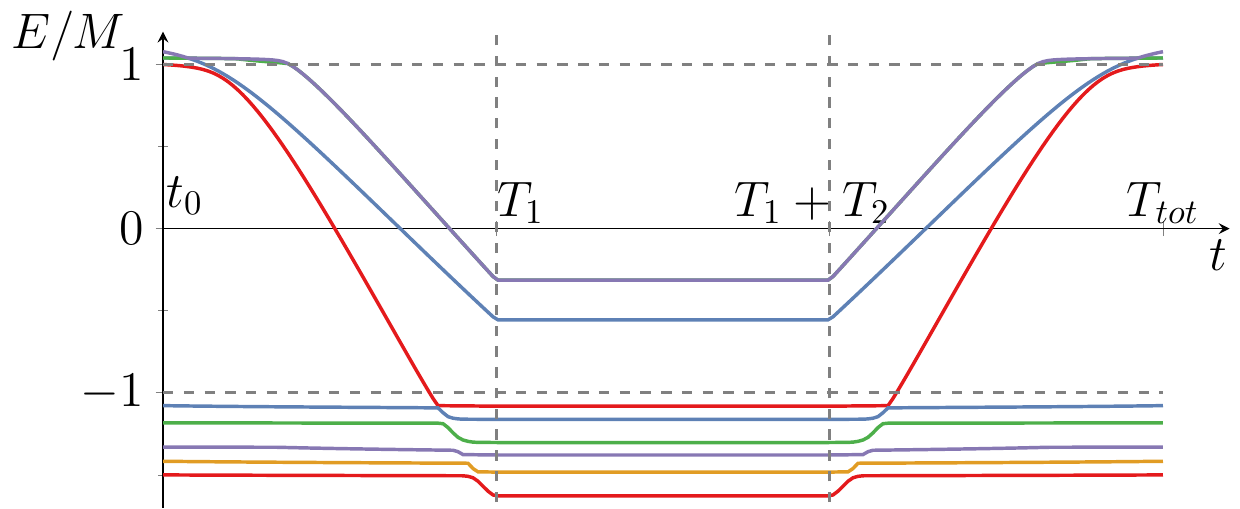}
  \caption{Top: Time--dependence of the potential's strength $\Phi(t, 0) = - \la(t) V_0$ 
    with three phases:
    switch-on, 
    static, 
    and switch-off.
    Bottom:
    energy values of bound states and discrete states in the discretized continuum 
    with anti--crossings during a supercritical process.
  }
  \label{fig:time-dep-V}
\end{figure}

In order to specify the time--dependence, 
we consider a family of time--dependent potentials 
$\Phi(t,\v x) = \Phi_{\la(t)}(\v x) = -\la(t) V(\v x)$, interpolating between the subcritical and supercritical regimes, cf. Fig. \ref{fig:time-dep-V}{, top}.
%
%
%
The quantum system described by the wave function $|\Psi(t)\>$ will evolve according to the Schrödinger equation
$ i\p_t |\Psi(t)\> = H(t) |\Psi(t)\> $
with time--dependent Hamiltonian $H(t)$. 
Its solution can be written by means of the evolution operator $U_t$ as
$ |\Psi(t)\> = U_t |\Psi(0)\> $.
Fig. \ref{fig:time-dep-V}{, bottom} presents an example of such a process. 
%
The avoided crossings 
enable an efficient transport of the amplitude of the dived bound state  into the negative pseudo-continuum and back. 

\xnewpage
\section{Spontaneous pair creation in optical lattice}

\subsection{Evolution in Fock space}

In the second quantized picture
we have, in the Heisenberg picture, 
$ \h \Psi(t, \v x) = \h U_t\, \h \Psi(0, \v x)\, \h U_t^\dagger$,
where $\h U_t$ is now a unitary evolution operator implemented in the many--particle Fock space $\F$ 
and $\h\Psi(t, \v x) = \sum_{p\in \Sigma_+} f_p(\v x)\, \h c_p(t) + \sum_{q\in \Sigma_-} g_q(\v x)\, \h d\s_q(t) $ is the time--dependent field operator
where $\Sigma_\pm$ are the positive and negative spectral subspaces of $H_0$ (for $\la = 0$).
$\h c_p$ and $\h d_q$ are particle and antiparticle (hole) annihilation operators
while $|f_p\>$ and $|g_q\>$ the corresponding basis wavefunctions
\cite{Scharf}.
We prepare the system to be initially in the ground state $|\Omega\>$ of the lattice half filled with fermions. 
It satisfies
$\h c_p(0) |\Omega \> = \h d_q(0) |\Omega\> = 0$
for all $p\in \Sigma_+$ and $q\in \Sigma_-$.
%
The instantaneous particle number operator 
$ \h N(t) = \sum_{p\in \Sigma_+} \h c\s_p(t) \h c_p(t) + \sum_{q\in \Sigma_-} \h d\s_q(t) \h d_q(t) $
gives initially $\< \Omega | \h N(0) | \Omega \> = 0$.
Since in the non--interacting theory the operator expectation values in the Fock space $\F$ can be reduced to expressions calculated in the single particle picture in the Hilbert space $\H$ \cite{Scharf},
we find 
%
\begin{equation} \label{eq:N-operator-HS}
  N(t) = \< \Omega | \h N(t) | \Omega \>  = || U(t)_{-+} ||^2_{HS} + || U(t)_{+-} ||^2_{HS}
\end{equation}
where $||...||_{HS}$ is the Hilbert--Schmidt norm, $U(t)_{\mp\pm} = P_\mp U(t) P_\pm$ and $P_\pm$ are projectors onto the initial positive/negative spectral subspaces of $H_0$ \cite{Thaller, NS-PhD}.
%

\subsection{Adiabatic limit for subcritical systems}

From the adiabatic theorem \cite{Teufel-AdiabaticTh} it follows that in subcritical systems, where no energy level crossings take place and the bound states stay away from the continuous spectrum, the tunneling between time--dependent eigenstates is suppressed as $T_{tot}$ increases. 
Accordingly, the total particle production rate vanishes 
in the adiabatic limit, as $T_{tot}\ra\infty$.
Our numerical simulations yield $N \sim 1/T_{tot}^\alpha \ra 0$ with $\alpha \approx 1.25$.


\subsection{Pseudo--adiabatic regime for supercritical systems}

The above holds no more when the system becomes temporarily supercritical. 
Then, the deepest bound state reaches the negative continuum, becomes a resonance and the adiabatic theorem breaks down \cite{Nen80, Nen87, Thaller, NS-PhD, NS-QED-JPhysA, Pickl+Duerr-AdiabSpPCr}.
A mixing between positive and negative energy states 
takes place and, accordingly, particle--antiparticle pairs are created.
In the adiabatic limit, the complex resonance $E_R$ with $\Re E_R < -M$
decays like $e^{- |\Im E_R| t}$ 
%
and that part of the wavefunction stays trapped in the negative continuum forever.
In a finite model with lattice size $\N \times \N$, the continuum is discretized with $\D E \sim \N^{-2}$ and there are no true resonances.
Instead, 
avoided crossings take place
with transition probabilities being significant when the duration of the processes $T \lesssim 1/\D E$ (cf. Landau--Zener model \cite{Zener}).
In consequence, the continuum limit and the adiabatic limit do not commute \cite{NS-PhD}. 
The correct limit to reproduce the spontaneous pair creation is thus
only possible in combination with the reduction of the discretization spacing $\D E$ and hence with scaling of the system size 
which is 
difficult to realize in practice. 
In contrast, for fixed system size, the time--scale is limited by $T < 1/\D E$ and
special effort is necessary to obtain results near the true adiabatic regime.
%
Results of such computations are presented in Fig. \ref{fig:Np-adiabatic-limit} and \ref{fig:Timedep_Pprod}.


\subsection{Spontaneous pair creation in supercritical fields}

\begin{figure}[ht]
  \includegraphics[width=0.7\linewidth]{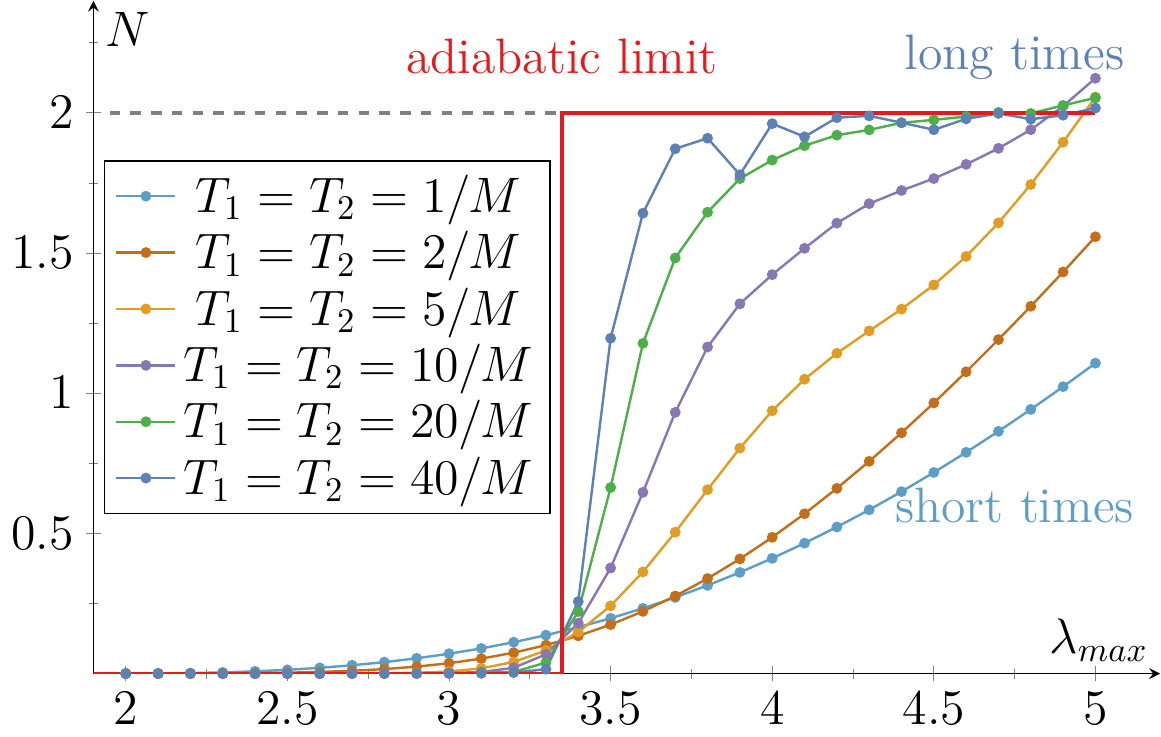}
  \caption{Total number of created particles and antiparticles $N$ as a function of maximal potential depth $\la_{max}$ for various total timescales approaching the theoretical adiabatic limit.}
  \label{fig:Np-adiabatic-limit}
\end{figure}

The total pair creation $N$ in time--dependent processes can be split into two contributions
$ N = N_{dyn} + N_{spont} $.
$N_{dyn}$ is the dynamical contribution related to time--dependence of the processes which vanishes in the adiabatic limit.
$N_{spont}$ is the spontaneous part which is solely related to the decay of the vacuum and, 
in the adiabatic limit, 
is zero for subcritical and 
twice the number of the dived states for supercritical processes (cf. Fig \ref{fig:Np-adiabatic-limit})
\begin{equation}
  N \ra \left\{ \begin{array}{ll}
                    0, & \la_{max} < \la_{cr} \\
                    2\, N_{dived} > 0, & \la_{max} > \la_{cr}
                 \end{array}
  \right. \quad \text{as} \quad t \ra \infty.
\end{equation}

\begin{figure}[ht]
  \includegraphics[width=0.9\linewidth]{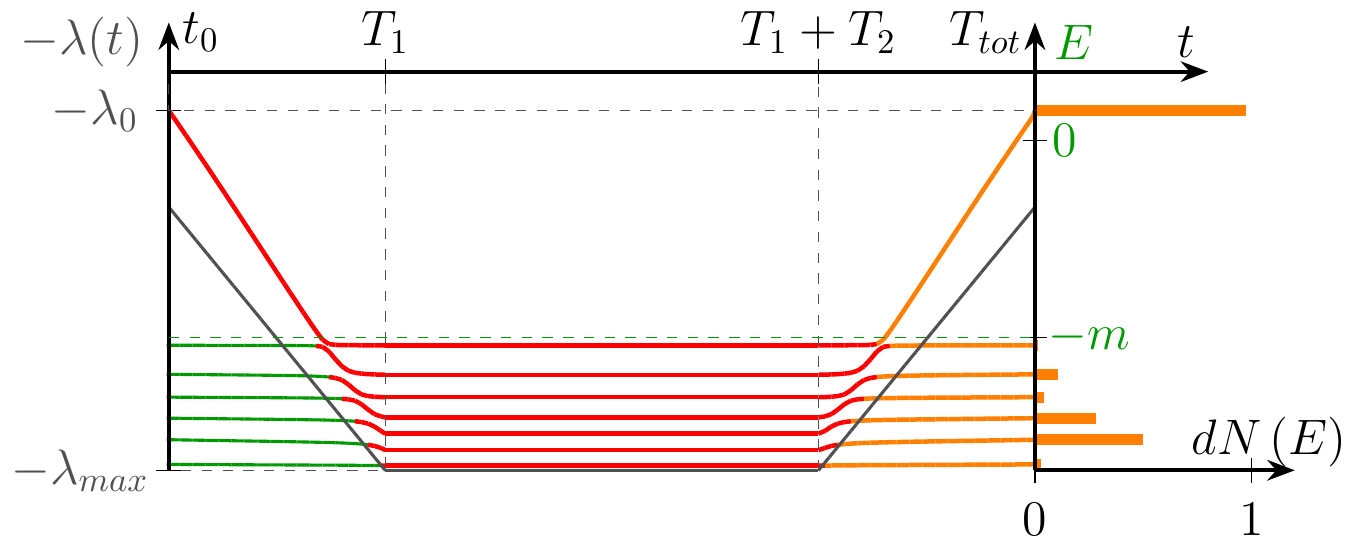}
  \caption{The course of the adiabatic time-evolution with $T_1 = T_2 = 20/M$. 
  The energy levels $E(t)$ are shown in the main $(t, E)$ plot (green and red solid lines) and particle production probabilities are shown in the right $(E, dN)$ plot (orange bars). 
  The potential's depth $-\la(t) V_0$ is overlaid (gray solid line, not to scale).
  The created antiparticles are peaked in the negative continuum around the resonance energy $E_R = -1.520 M$.
  The particle is created in the bound state at $E_{bound} = 0.153 M$.
  }
  \label{fig:Timedep_Pprod}
\end{figure}

The changeover of the total pair creation rate $N$ in the adiabatic regime between sub- and supercritical processes is qualitative and, in the exact adiabatic limit, discontinuous \cite{Nen80, Nen87, Thaller, NS-PhD, NS-QED-JPhysA, Pickl+Duerr-AdiabSpPCr}.
Fig. \ref{fig:Np-adiabatic-limit} shows its realization in the lattice system with an external time--dependent potential $\Phi(t,\v x) = -\la(t) V(\v x)$.
For subcritical $\la_{max}<\la_{cr}$ the particle creation clearly tends to zero (adiabatic limit) while for supercritical $\la_{max}>\la_{cr}$ the number slowly reaches $2$ which corresponds to one particle and one anti--particle (cf. Eq. \eqref{eq:N-operator-HS}) for one dived state. For values $\la \approx 5.4 > \la_{cr}$ another bound state dives into the negative continuum and $N$ increases to $4$.
Fig.  \ref{fig:Timedep_Pprod} shows a typical supercritical process together with the final energy distribution of created particles and antiparticles. 

\section{Optical lattice setup}

The realization of the  proposed lattice Hamiltonian \eqref{eq:discr-Ham} requires two types of sites with complex tunneling amplitudes $t_{n,m,d} = e^{i \pi [(n+m)(d-1) + d/2]}$ for lattice sites $(n,m)$ and hopping directions $d=0,1,2,3$.
It can be realized with ultra--cold fermionic atoms in bichromatic square optical lattices 
utilizing laser assisted hopping 
\cite{Lewenstein-Dirac-CurvedST, Unruh+OL-Tarruell+Celi+Lewenstein,
LaserAssistedHoppingOptLat, MasslessDirac-SqLattice, Dirac-StaggeredMagnField, Lewenstein-Dirac-CurvedST, QED-In-OptLat, FieldTheo-QSim-In-OptLat, StrongB-Staggered-OptLat}.
%
An alternative setup with real tunneling constants, which might be easier to realize, can be based on a hexagonal lattice. Its analysis will be published elsewhere.
%
%
The lattice size $51 \times 51$ used in the calculations was motivated by the experimental possibilities \cite{Bloch+Dalibard-OptLat-Review}.
The external potential can be created by a perpendicular laser beam \cite{Bloch-OptLat}.


For more detailed experimental protocol, describing the preparation of the initial state, evolution and measurement, we refer to our previous works \cite{NS+RS-BiOptLat-Lett, NS+RS-BiOptLat}. 
It should utilize the techniques of coherent transfer between lattice bands and band spectroscopy \cite{BEC-OptLat_BandTransport}.

\section{Discussion}

The proposed quantum simulator mimics the quantum many--particle
system consisting of electrons and positrons in presence of a strong electric field
and should thereby reproduce the effect of spontaneous pair creation
which has not yet been demonstrated within QED. 
%
The proposed analogue model for the spontaneous pair creation can be realized in a table--top experiment.
The setup is experimentally challenging yet feasible \cite{Lewenstein-Dirac-CurvedST, Unruh+OL-Tarruell+Celi+Lewenstein, LaserAssistedHoppingOptLat}.
Analog quantum simulations seem very valuable for a better understanding of the fundamental principles of quantum field theory under extreme conditions.
The proposed one will facilitate investigation of space--time dependent electric fields 
between the adiabatic and quench regimes. 
It can also provide new insight into the role of interactions which may be incorporated into the simulator.
The latter can offer a foundation for study of similar phenomena in solid states, e.g. in graphene \cite{Pereira+CastroNeto-Supercritical-Graphene, Katsnelson+Levitov-VacuumPol-Supercritical-Graphene}. 

\section{Acknowledgments}

We gratefully acknowledge the funding by the Deutsche Forschungs\-gemeinschaft (DFG, German Research Foundation) --
Project 278162697 -- SFB 1242.


\bibliography{lattices}

\begin{thebibliography}{39}
\expandafter\ifx\csname natexlab\endcsname\relax\def\natexlab#1{#1}\fi
\expandafter\ifx\csname bibnamefont\endcsname\relax
  \def\bibnamefont#1{#1}\fi
\expandafter\ifx\csname bibfnamefont\endcsname\relax
  \def\bibfnamefont#1{#1}\fi
\expandafter\ifx\csname citenamefont\endcsname\relax
  \def\citenamefont#1{#1}\fi
\expandafter\ifx\csname url\endcsname\relax
  \def\url#1{\texttt{#1}}\fi
\expandafter\ifx\csname urlprefix\endcsname\relax\def\urlprefix{URL }\fi
\providecommand{\bibinfo}[2]{#2}
\providecommand{\eprint}[2][]{\url{#2}}

\bibitem[{\citenamefont{Thaller}(1992)}]{Thaller}
\bibinfo{author}{\bibfnamefont{B.}~\bibnamefont{Thaller}},
  \emph{\bibinfo{title}{The {D}irac Equation}} (\bibinfo{publisher}{Springer
  Verlag Berlin Heidelberg}, \bibinfo{year}{1992}).

\bibitem[{\citenamefont{Rafelski et~al.}(1974)\citenamefont{Rafelski,
  M{\"u}ller, and Greiner}}]{Rafelski+Mueller+Greiner-ChargedVac-Overcrit}
\bibinfo{author}{\bibfnamefont{J.}~\bibnamefont{Rafelski}},
  \bibinfo{author}{\bibfnamefont{B.}~\bibnamefont{M{\"u}ller}},
  \bibnamefont{and} \bibinfo{author}{\bibfnamefont{W.}~\bibnamefont{Greiner}},
  \bibinfo{journal}{Nuclear Physics B} \textbf{\bibinfo{volume}{68}},
  \bibinfo{pages}{585} (\bibinfo{year}{1974}).

\bibitem[{\citenamefont{Greiner et~al.}(1985)\citenamefont{Greiner, M{\"u}ller,
  and Rafelski}}]{GrMuRa}
\bibinfo{author}{\bibfnamefont{W.}~\bibnamefont{Greiner}},
  \bibinfo{author}{\bibfnamefont{B.}~\bibnamefont{M{\"u}ller}},
  \bibnamefont{and} \bibinfo{author}{\bibfnamefont{J.}~\bibnamefont{Rafelski}},
  \emph{\bibinfo{title}{Quantum Electrodynamics of Strong Fields}}, Texts and
  Monographs in Physics (\bibinfo{publisher}{Springer-Verlag},
  \bibinfo{year}{1985}).

\bibitem[{\citenamefont{Szpak}(2006)}]{NS-PhD}
\bibinfo{author}{\bibfnamefont{N.}~\bibnamefont{Szpak}}, Ph.D. thesis,
  \bibinfo{school}{University, Frankfurt am Main, Germany}
  (\bibinfo{year}{2006}).

\bibitem[{\citenamefont{Szpak}(2008)}]{NS-QED-JPhysA}
\bibinfo{author}{\bibfnamefont{N.}~\bibnamefont{Szpak}}, \bibinfo{journal}{J.
  Phys. A: Math. Theor.} \textbf{\bibinfo{volume}{41}}, \bibinfo{pages}{164059}
  (\bibinfo{year}{2008}).

\bibitem[{\citenamefont{Pickl and D{\"u}rr}(2008)}]{Pickl+Duerr-AdiabSpPCr}
\bibinfo{author}{\bibfnamefont{P.}~\bibnamefont{Pickl}} \bibnamefont{and}
  \bibinfo{author}{\bibfnamefont{D.}~\bibnamefont{D{\"u}rr}},
  \bibinfo{journal}{EPL (Europhysics Letters)} \textbf{\bibinfo{volume}{81}},
  \bibinfo{pages}{40001} (\bibinfo{year}{2008}).

\bibitem[{\citenamefont{Rafelski et~al.}(2017)\citenamefont{Rafelski, Kirsch,
  M{\"u}ller, Reinhardt, and Greiner}}]{Rafelski2017-ProbingVacuum}
\bibinfo{author}{\bibfnamefont{J.}~\bibnamefont{Rafelski}},
  \bibinfo{author}{\bibfnamefont{J.}~\bibnamefont{Kirsch}},
  \bibinfo{author}{\bibfnamefont{B.}~\bibnamefont{M{\"u}ller}},
  \bibinfo{author}{\bibfnamefont{J.}~\bibnamefont{Reinhardt}},
  \bibnamefont{and} \bibinfo{author}{\bibfnamefont{W.}~\bibnamefont{Greiner}},
  in \emph{\bibinfo{booktitle}{New horizons in fundamental physics}}
  (\bibinfo{publisher}{Springer}, \bibinfo{year}{2017}), pp.
  \bibinfo{pages}{211--251}.

\bibitem[{\citenamefont{Bloch}(2005)}]{Bloch-OptLat}
\bibinfo{author}{\bibfnamefont{I.}~\bibnamefont{Bloch}},
  \bibinfo{journal}{Nature Physics} \textbf{\bibinfo{volume}{1}},
  \bibinfo{pages}{23} (\bibinfo{year}{2005}).

\bibitem[{\citenamefont{Greiner et~al.}(2002)\citenamefont{Greiner, Mandel,
  Esslinger, H{\"a}nsch, and
  Bloch}}]{MGreiner+Esslinger+Haensch+Bloch-SuperfluidMottOptLat}
\bibinfo{author}{\bibfnamefont{M.}~\bibnamefont{Greiner}},
  \bibinfo{author}{\bibfnamefont{O.}~\bibnamefont{Mandel}},
  \bibinfo{author}{\bibfnamefont{T.}~\bibnamefont{Esslinger}},
  \bibinfo{author}{\bibfnamefont{T.~W.} \bibnamefont{H{\"a}nsch}},
  \bibnamefont{and} \bibinfo{author}{\bibfnamefont{I.}~\bibnamefont{Bloch}},
  \bibinfo{journal}{Nature} \textbf{\bibinfo{volume}{415}}, \bibinfo{pages}{39}
  (\bibinfo{year}{2002}).

\bibitem[{\citenamefont{Lewenstein et~al.}(2007)\citenamefont{Lewenstein,
  Sanpera, Ahufinger, Damski, Sen, and
  Sen}}]{Damski+Lewenstein+Sanpera-OpticalLattices-Review}
\bibinfo{author}{\bibfnamefont{M.}~\bibnamefont{Lewenstein}},
  \bibinfo{author}{\bibfnamefont{A.}~\bibnamefont{Sanpera}},
  \bibinfo{author}{\bibfnamefont{V.}~\bibnamefont{Ahufinger}},
  \bibinfo{author}{\bibfnamefont{B.}~\bibnamefont{Damski}},
  \bibinfo{author}{\bibfnamefont{A.}~\bibnamefont{Sen}}, \bibnamefont{and}
  \bibinfo{author}{\bibfnamefont{U.}~\bibnamefont{Sen}},
  \bibinfo{journal}{Advances in Physics} \textbf{\bibinfo{volume}{56}},
  \bibinfo{pages}{243} (\bibinfo{year}{2007}).

\bibitem[{\citenamefont{Esslinger}(2010)}]{Esslinger-FermiHubbard-Review}
\bibinfo{author}{\bibfnamefont{T.}~\bibnamefont{Esslinger}},
  \bibinfo{journal}{Annu. Rev. Condens. Matter Phys.}
  \textbf{\bibinfo{volume}{1}}, \bibinfo{pages}{129} (\bibinfo{year}{2010}).

\bibitem[{\citenamefont{Krutitsky}(2016)}]{Krutitsky-BoseHubbard-Review}
\bibinfo{author}{\bibfnamefont{K.~V.} \bibnamefont{Krutitsky}},
  \bibinfo{journal}{Physics Reports} \textbf{\bibinfo{volume}{607}},
  \bibinfo{pages}{1} (\bibinfo{year}{2016}).

\bibitem[{\citenamefont{Bloch et~al.}(2008)\citenamefont{Bloch, Dalibard, and
  Zwerger}}]{Bloch+Dalibard-OptLat-Review}
\bibinfo{author}{\bibfnamefont{I.}~\bibnamefont{Bloch}},
  \bibinfo{author}{\bibfnamefont{J.}~\bibnamefont{Dalibard}}, \bibnamefont{and}
  \bibinfo{author}{\bibfnamefont{W.}~\bibnamefont{Zwerger}},
  \bibinfo{journal}{Reviews of Modern Physics} \textbf{\bibinfo{volume}{80}},
  \bibinfo{pages}{885} (\bibinfo{year}{2008}).

\bibitem[{\citenamefont{Yukalov}(2009)}]{Yukalov-OptLat-Review}
\bibinfo{author}{\bibfnamefont{V.}~\bibnamefont{Yukalov}},
  \bibinfo{journal}{Laser Physics} \textbf{\bibinfo{volume}{19}},
  \bibinfo{pages}{1} (\bibinfo{year}{2009}).

\bibitem[{\citenamefont{Pereira et~al.}(2008)\citenamefont{Pereira, Kotov, and
  Neto}}]{Pereira+CastroNeto-Supercritical-Graphene}
\bibinfo{author}{\bibfnamefont{V.~M.} \bibnamefont{Pereira}},
  \bibinfo{author}{\bibfnamefont{V.~N.} \bibnamefont{Kotov}}, \bibnamefont{and}
  \bibinfo{author}{\bibfnamefont{A.~C.} \bibnamefont{Neto}},
  \bibinfo{journal}{Phys. Rev. B} \textbf{\bibinfo{volume}{78}},
  \bibinfo{pages}{085101} (\bibinfo{year}{2008}).

\bibitem[{\citenamefont{Shytov et~al.}(2007)\citenamefont{Shytov, Katsnelson,
  and Levitov}}]{Katsnelson+Levitov-VacuumPol-Supercritical-Graphene}
\bibinfo{author}{\bibfnamefont{A.~V.} \bibnamefont{Shytov}},
  \bibinfo{author}{\bibfnamefont{M.~I.} \bibnamefont{Katsnelson}},
  \bibnamefont{and} \bibinfo{author}{\bibfnamefont{L.~S.}
  \bibnamefont{Levitov}}, \bibinfo{journal}{Phys. Rev. Lett.}
  \textbf{\bibinfo{volume}{99}}, \bibinfo{pages}{236801}
  (\bibinfo{year}{2007}).

\bibitem[{\citenamefont{Longhi}(2010)}]{Dirac-photons}
\bibinfo{author}{\bibfnamefont{S.}~\bibnamefont{Longhi}},
  \bibinfo{journal}{Phys. Rev. A} \textbf{\bibinfo{volume}{81}},
  \bibinfo{pages}{022118} (\bibinfo{year}{2010}).

\bibitem[{\citenamefont{Witthaut et~al.}(2011)\citenamefont{Witthaut, Salger,
  Kling, Grossert, and Weitz}}]{Witthaut+Weitz-DoublePeriodicOptPot}
\bibinfo{author}{\bibfnamefont{D.}~\bibnamefont{Witthaut}},
  \bibinfo{author}{\bibfnamefont{T.}~\bibnamefont{Salger}},
  \bibinfo{author}{\bibfnamefont{S.}~\bibnamefont{Kling}},
  \bibinfo{author}{\bibfnamefont{C.}~\bibnamefont{Grossert}}, \bibnamefont{and}
  \bibinfo{author}{\bibfnamefont{M.}~\bibnamefont{Weitz}},
  \bibinfo{journal}{Phys. Rev. A} \textbf{\bibinfo{volume}{84}},
  \bibinfo{pages}{033601} (\bibinfo{year}{2011}).

\bibitem[{\citenamefont{Zhu et~al.}(2007)\citenamefont{Zhu, Wang, and
  Duan}}]{QSim-Dirac-HexLattice}
\bibinfo{author}{\bibfnamefont{S.-L.} \bibnamefont{Zhu}},
  \bibinfo{author}{\bibfnamefont{B.}~\bibnamefont{Wang}}, \bibnamefont{and}
  \bibinfo{author}{\bibfnamefont{L.-M.} \bibnamefont{Duan}},
  \bibinfo{journal}{Phys. Rev. Lett.} \textbf{\bibinfo{volume}{98}},
  \bibinfo{pages}{260402} (\bibinfo{year}{2007}).

\bibitem[{\citenamefont{Cirac et~al.}(2010)\citenamefont{Cirac, Maraner, and
  Pachos}}]{Dirac+Interaction-OptLat}
\bibinfo{author}{\bibfnamefont{J.~I.} \bibnamefont{Cirac}},
  \bibinfo{author}{\bibfnamefont{P.}~\bibnamefont{Maraner}}, \bibnamefont{and}
  \bibinfo{author}{\bibfnamefont{J.~K.} \bibnamefont{Pachos}},
  \bibinfo{journal}{Phys. Rev. Lett.} \textbf{\bibinfo{volume}{105}},
  \bibinfo{pages}{190403} (\bibinfo{year}{2010}).

\bibitem[{\citenamefont{Hou et~al.}(2009)\citenamefont{Hou, Yang, and
  Liu}}]{MasslessDirac-SqLattice}
\bibinfo{author}{\bibfnamefont{J.-M.} \bibnamefont{Hou}},
  \bibinfo{author}{\bibfnamefont{W.-X.} \bibnamefont{Yang}}, \bibnamefont{and}
  \bibinfo{author}{\bibfnamefont{X.-J.} \bibnamefont{Liu}},
  \bibinfo{journal}{Phys. Rev. A} \textbf{\bibinfo{volume}{79}},
  \bibinfo{pages}{043621} (\bibinfo{year}{2009}).

\bibitem[{\citenamefont{Lim et~al.}(2008)\citenamefont{Lim, Smith, and
  Hemmerich}}]{Dirac-StaggeredMagnField}
\bibinfo{author}{\bibfnamefont{L.-K.} \bibnamefont{Lim}},
  \bibinfo{author}{\bibfnamefont{C.~M.} \bibnamefont{Smith}}, \bibnamefont{and}
  \bibinfo{author}{\bibfnamefont{A.}~\bibnamefont{Hemmerich}},
  \bibinfo{journal}{Phys. Rev. Lett.} \textbf{\bibinfo{volume}{100}},
  \bibinfo{pages}{130402} (\bibinfo{year}{2008}).

\bibitem[{\citenamefont{Goldman et~al.}(2009)\citenamefont{Goldman, Kubasiak,
  Bermudez, Gaspard, Lewenstein, and
  Martin-Delgado}}]{Goldman+Lewenstein-Dirac-by-SU2}
\bibinfo{author}{\bibfnamefont{N.}~\bibnamefont{Goldman}},
  \bibinfo{author}{\bibfnamefont{A.}~\bibnamefont{Kubasiak}},
  \bibinfo{author}{\bibfnamefont{A.}~\bibnamefont{Bermudez}},
  \bibinfo{author}{\bibfnamefont{P.}~\bibnamefont{Gaspard}},
  \bibinfo{author}{\bibfnamefont{M.}~\bibnamefont{Lewenstein}},
  \bibnamefont{and} \bibinfo{author}{\bibfnamefont{M.~A.}
  \bibnamefont{Martin-Delgado}}, \bibinfo{journal}{Phys. Rev. Lett.}
  \textbf{\bibinfo{volume}{103}}, \bibinfo{pages}{035301}
  (\bibinfo{year}{2009}).

\bibitem[{\citenamefont{Boada et~al.}(2011)\citenamefont{Boada, Celi, Latorre,
  and Lewenstein}}]{Lewenstein-Dirac-CurvedST}
\bibinfo{author}{\bibfnamefont{O.}~\bibnamefont{Boada}},
  \bibinfo{author}{\bibfnamefont{A.}~\bibnamefont{Celi}},
  \bibinfo{author}{\bibfnamefont{J.}~\bibnamefont{Latorre}}, \bibnamefont{and}
  \bibinfo{author}{\bibfnamefont{M.}~\bibnamefont{Lewenstein}},
  \bibinfo{journal}{New Journal of Physics} \textbf{\bibinfo{volume}{13}},
  \bibinfo{pages}{035002} (\bibinfo{year}{2011}).

\bibitem[{\citenamefont{Mazza et~al.}(2012)\citenamefont{Mazza, Bermudez,
  Goldman, Rizzi, Martin-Delgado, and Lewenstein}}]{FieldTheo-QSim-In-OptLat}
\bibinfo{author}{\bibfnamefont{L.}~\bibnamefont{Mazza}},
  \bibinfo{author}{\bibfnamefont{A.}~\bibnamefont{Bermudez}},
  \bibinfo{author}{\bibfnamefont{N.}~\bibnamefont{Goldman}},
  \bibinfo{author}{\bibfnamefont{M.}~\bibnamefont{Rizzi}},
  \bibinfo{author}{\bibfnamefont{M.~A.} \bibnamefont{Martin-Delgado}},
  \bibnamefont{and}
  \bibinfo{author}{\bibfnamefont{M.}~\bibnamefont{Lewenstein}},
  \bibinfo{journal}{New Journal of Physics} \textbf{\bibinfo{volume}{14}},
  \bibinfo{pages}{015007} (\bibinfo{year}{2012}).

\bibitem[{\citenamefont{Aidelsburger et~al.}(2011)\citenamefont{Aidelsburger,
  Atala, Nascimb{\`e}ne, Trotzky, Chen, and Bloch}}]{StrongB-Staggered-OptLat}
\bibinfo{author}{\bibfnamefont{M.}~\bibnamefont{Aidelsburger}},
  \bibinfo{author}{\bibfnamefont{M.}~\bibnamefont{Atala}},
  \bibinfo{author}{\bibfnamefont{S.}~\bibnamefont{Nascimb{\`e}ne}},
  \bibinfo{author}{\bibfnamefont{S.}~\bibnamefont{Trotzky}},
  \bibinfo{author}{\bibfnamefont{Y.-A.} \bibnamefont{Chen}}, \bibnamefont{and}
  \bibinfo{author}{\bibfnamefont{I.}~\bibnamefont{Bloch}},
  \bibinfo{journal}{Phys. Rev. Lett.} \textbf{\bibinfo{volume}{107}},
  \bibinfo{pages}{255301} (\bibinfo{year}{2011}).

\bibitem[{\citenamefont{Szpak and Sch{\"u}tzhold}(2011)}]{NS+RS-BiOptLat-Lett}
\bibinfo{author}{\bibfnamefont{N.}~\bibnamefont{Szpak}} \bibnamefont{and}
  \bibinfo{author}{\bibfnamefont{R.}~\bibnamefont{Sch{\"u}tzhold}},
  \bibinfo{journal}{Phys. Rev. A} \textbf{\bibinfo{volume}{84}},
  \bibinfo{pages}{050101} (\bibinfo{year}{2011}).

\bibitem[{\citenamefont{Szpak and Sch{\"u}tzhold}(2012)}]{NS+RS-BiOptLat}
\bibinfo{author}{\bibfnamefont{N.}~\bibnamefont{Szpak}} \bibnamefont{and}
  \bibinfo{author}{\bibfnamefont{R.}~\bibnamefont{Sch{\"u}tzhold}},
  \bibinfo{journal}{New Journal of Physics} \textbf{\bibinfo{volume}{14}},
  \bibinfo{pages}{035001} (\bibinfo{year}{2012}).

\bibitem[{\citenamefont{Salger et~al.}(2007)\citenamefont{Salger, Geckeler,
  Kling, and Weitz}}]{Weitz-DoublePeriodicOptPot}
\bibinfo{author}{\bibfnamefont{T.}~\bibnamefont{Salger}},
  \bibinfo{author}{\bibfnamefont{C.}~\bibnamefont{Geckeler}},
  \bibinfo{author}{\bibfnamefont{S.}~\bibnamefont{Kling}}, \bibnamefont{and}
  \bibinfo{author}{\bibfnamefont{M.}~\bibnamefont{Weitz}},
  \bibinfo{journal}{Phys. Rev. Lett.} \textbf{\bibinfo{volume}{99}},
  \bibinfo{pages}{190405} (\bibinfo{year}{2007}).

\bibitem[{\citenamefont{Susskind}(1977)}]{Susskind-FermionDoubling}
\bibinfo{author}{\bibfnamefont{L.}~\bibnamefont{Susskind}},
  \bibinfo{journal}{Phys. Rev. D} \textbf{\bibinfo{volume}{16}},
  \bibinfo{pages}{3031} (\bibinfo{year}{1977}).

\bibitem[{\citenamefont{Scharf}(1995)}]{Scharf}
\bibinfo{author}{\bibfnamefont{G.}~\bibnamefont{Scharf}},
  \emph{\bibinfo{title}{Finite Quantum Electrodynamics - The Causal Approach,
  2nd Edition}} (\bibinfo{publisher}{Springer Verlag Berlin Heidelberg New
  York}, \bibinfo{year}{1995}).

\bibitem[{\citenamefont{Teufel}(2003)}]{Teufel-AdiabaticTh}
\bibinfo{author}{\bibfnamefont{S.}~\bibnamefont{Teufel}},
  \emph{\bibinfo{title}{Adiabatic perturbation theory in quantum dynamics}}
  (\bibinfo{publisher}{Springer}, \bibinfo{year}{2003}).

\bibitem[{\citenamefont{Nenciu}(1980)}]{Nen80}
\bibinfo{author}{\bibfnamefont{G.}~\bibnamefont{Nenciu}},
  \bibinfo{journal}{Comm. Math. Phys.} \textbf{\bibinfo{volume}{76}},
  \bibinfo{pages}{117} (\bibinfo{year}{1980}).

\bibitem[{\citenamefont{Nenciu}(1987)}]{Nen87}
\bibinfo{author}{\bibfnamefont{G.}~\bibnamefont{Nenciu}},
  \bibinfo{journal}{Comm. Math. Phys.} \textbf{\bibinfo{volume}{109}},
  \bibinfo{pages}{303} (\bibinfo{year}{1987}).

\bibitem[{\citenamefont{Zener}(1932)}]{Zener}
\bibinfo{author}{\bibfnamefont{C.}~\bibnamefont{Zener}},
  \bibinfo{journal}{Proc. R. Soc. Lond. A} \textbf{\bibinfo{volume}{137}},
  \bibinfo{pages}{696} (\bibinfo{year}{1932}).

\bibitem[{\citenamefont{Rodr{\'\i}guez-Laguna
  et~al.}(2017)\citenamefont{Rodr{\'\i}guez-Laguna, Tarruell, Lewenstein, and
  Celi}}]{Unruh+OL-Tarruell+Celi+Lewenstein}
\bibinfo{author}{\bibfnamefont{J.}~\bibnamefont{Rodr{\'\i}guez-Laguna}},
  \bibinfo{author}{\bibfnamefont{L.}~\bibnamefont{Tarruell}},
  \bibinfo{author}{\bibfnamefont{M.}~\bibnamefont{Lewenstein}},
  \bibnamefont{and} \bibinfo{author}{\bibfnamefont{A.}~\bibnamefont{Celi}},
  \bibinfo{journal}{Phys. Rev. A} \textbf{\bibinfo{volume}{95}},
  \bibinfo{pages}{013627} (\bibinfo{year}{2017}).

\bibitem[{\citenamefont{Miyake et~al.}(2013)\citenamefont{Miyake, Siviloglou,
  Kennedy, Burton, and Ketterle}}]{LaserAssistedHoppingOptLat}
\bibinfo{author}{\bibfnamefont{H.}~\bibnamefont{Miyake}},
  \bibinfo{author}{\bibfnamefont{G.~A.} \bibnamefont{Siviloglou}},
  \bibinfo{author}{\bibfnamefont{C.~J.} \bibnamefont{Kennedy}},
  \bibinfo{author}{\bibfnamefont{W.~C.} \bibnamefont{Burton}},
  \bibnamefont{and} \bibinfo{author}{\bibfnamefont{W.}~\bibnamefont{Ketterle}},
  \bibinfo{journal}{Phys. Rev. Lett.} \textbf{\bibinfo{volume}{111}},
  \bibinfo{pages}{185302} (\bibinfo{year}{2013}).

\bibitem[{\citenamefont{Kapit and Mueller}(2011)}]{QED-In-OptLat}
\bibinfo{author}{\bibfnamefont{E.}~\bibnamefont{Kapit}} \bibnamefont{and}
  \bibinfo{author}{\bibfnamefont{E.}~\bibnamefont{Mueller}},
  \bibinfo{journal}{Phys. Rev. A} \textbf{\bibinfo{volume}{83}},
  \bibinfo{pages}{033625} (\bibinfo{year}{2011}).

\bibitem[{\citenamefont{Denschlag et~al.}(2002)\citenamefont{Denschlag,
  Simsarian, H{\"a}ffner, McKenzie, Browaeys, Cho, Helmerson, Rolston, and
  Phillips}}]{BEC-OptLat_BandTransport}
\bibinfo{author}{\bibfnamefont{J.~H.} \bibnamefont{Denschlag}},
  \bibinfo{author}{\bibfnamefont{J.~E.} \bibnamefont{Simsarian}},
  \bibinfo{author}{\bibfnamefont{H.}~\bibnamefont{H{\"a}ffner}},
  \bibinfo{author}{\bibfnamefont{C.}~\bibnamefont{McKenzie}},
  \bibinfo{author}{\bibfnamefont{A.}~\bibnamefont{Browaeys}},
  \bibinfo{author}{\bibfnamefont{D.}~\bibnamefont{Cho}},
  \bibinfo{author}{\bibfnamefont{K.}~\bibnamefont{Helmerson}},
  \bibinfo{author}{\bibfnamefont{S.~L.} \bibnamefont{Rolston}},
  \bibnamefont{and} \bibinfo{author}{\bibfnamefont{W.~D.}
  \bibnamefont{Phillips}}, \bibinfo{journal}{Journal of Physics B: Atomic,
  Molecular and Optical Physics} \textbf{\bibinfo{volume}{35}},
  \bibinfo{pages}{3095} (\bibinfo{year}{2002}).

\end{thebibliography}
\bibliographystyle{apsrev}


\end{document}